\documentclass[12pt]{article}
\usepackage{amstex}

\begin{document}
\title{String Cosmology}
\author{John D. Barrow and Kerstin E. Kunze\\
Astronomy Centre\\
University of Sussex\\
Brighton BN1 9QJ\\
UK}
\date{}
\maketitle

\begin{abstract}
An overview is given of the formulation of low-energy string cosmologies
together with examples of particular solutions, successes and problems of
the theory.
\end{abstract}

\section{Introduction}

General relativity provides a successful theory to describe classical
phenomena in gravity. Several experiments provide impressive evidence for
how well it succeeds. The Hulse-Taylor binary pulsar PSR1913+16 consists of
a pulsating and a non-pulsating neutron star. According to general
relativity the orbit should slowly decay and the orbital period should
increase due to the loss of energy by the emission of gravitational waves.
This has been indeed observed \cite{ref1a} to very high precision. However,
despite this outstanding success,  under certain extreme physical conditions
general relativity predicts that it cannot predict \cite{refA}. We expect
general relativity to be a low-energy and low space-time curvature limit of
some quantum theory of gravitation that has yet to be found. String
cosmology is one approach to discover the cosmological consequences of some
of these quantum gravitational corrections to the general theory of
relativity.

The search for a unification of quantum field theory and general relativity
is the most important quest in theoretical physics. Unfortunately, the
canonical quantum field approach to quantum gravity is riddled with
incurable infinities and so is not well defined. Superstring theory, in
contrast, is a very promising candidate for a consistent theory of quantum
gravity \cite{ref1}. In all, there are five consistent superstring theories,
namely, type $I,IIA,IIB,$ $E_8\times E_8$ heterotic and $SO(32)$ heterotic,
which all seem to be related by duality symmetries facilitating the
existence of a more fundamental underlying theory. Attempts to identify this
underlying M-theory (see, for example, \cite{ref3a}) have yet to arrive at a
complete final formulation and interpretation.

Due to anomaly considerations, superstring theories are only consistent in
ten dimensional space-times. In order to make contact with the observed
four-dimensional world it is assumed that the extra six spatial dimensions
are compactified. For simplicity, it is often assumed that the internal
space is a torus. However, particle physics requires the internal space to
be more complicated, namely of Calabi-Yau type \cite{ref53}. In their
low-energy limit string theories predict a classical gravitational
interaction. This theory of gravity differs from standard Einstein
relativity. However, general relativity is a particular solution. In the
simplest models, and to lowest order, the corrections to standard gravity
come in the form of two fields: the dilaton, which is a scalar field, and
the antisymmetric tensor field strength, which is a three-form (see, for
example, \cite{ref49}). In principle, moduli fields from the compactified
dimensions should also be taken into account. In four dimensions, the
antisymmetric tensor field strength can be described in terms of its (Hodge)
dual which is the gradient of a scalar field. So, in the simplest models the
gravitational sector of low-energy string theory describes general
relativity coupled to two stiff perfect fluids.

Most approaches to string cosmology begin with the low-energy effective
action. However, as in general relativity, the low-energy theory suffers
from space-time curvature singularities. An interesting example of a
regular, inhomogeneous solution is given in \cite{refG}. In regions of high
curvature and/or string coupling the low-energy limit of string theory can
no longer be trusted and higher-order corrections have to be taken into
account \cite{high}. However, it is an open question, as to when a solution
to the low-energy theory is also an exact solution to all orders or can be
extended to one. A notable exception is provided by plane-wave space-times,
which are exact solutions to string theory \cite{plane}. Other exact
solutions corresponding to conformal field theories with a cosmological
interpretation are discussed for example in \cite{exact}.

The inflationary paradigm \cite{ref55} has become an important extension of
the standard big bang model in order to explain certain cosmological puzzles
and provide seeds for large-scale structure formation. The implementation of
inflation in the low-energy limit of string theory requires some thought
since in the simplest models there is no potential for the dilaton, which
would be needed for the standard inflationary picture where the evolution of
the inflaton is driven by a non-vanishing potential energy density \cite
{ref35}. Any potential (for example, arising from supersymmetry breaking) is
too steep to dominate the expansion dynamics and leaves the kinetic energy
of the dilaton to dominate \cite{ref38}. One possibility to implement
inflation in string theory was discussed in \cite{infl}.

Another interesting approach is suggested by the pre-big-bang model \cite
{ppb}. Ten-dimensional Minkowski space-time is an exact solution to the full
superstring theory. This corresponds to a free theory where the string
coupling vanishes ($g\rightarrow 0$) as the dilaton goes off to minus
infinity ($\Phi \rightarrow -\infty $). The basic assumption of pre-big-bang
string cosmologies is that the evolution of the universe did indeed begin
near this trivial but exact vacuum of string theory \cite{ref47}.

\section{Low-energy Effective String Theory}

The low-energy effective action compactified to four dimensions is given by 
\cite{ref49} \cite{ref4} 
\begin{eqnarray}
S=\frac 1{2\ }\int d^4x\sqrt{-g}e^{-\Phi }\left[ R+\nabla _\mu \Phi \nabla
^\mu \Phi -\frac 1{12}H_{\mu \nu \rho }H^{\mu \nu \rho }\right]
\label{act1.7}
\end{eqnarray}
where $\Phi $ is the dilaton, $H_{\mu \nu \lambda }$ is the antisymmetric
tensor field strength which is defined by 
\begin{equation*}
H_{\mu \nu \rho }=\partial _{[\mu }B_{\nu \rho ]}.
\end{equation*}

In general, the low-energy effective action can be written in the form of
the Einstein-Hilbert action of general relativity in 4 dimensions by
applying the conformal transformation of the space-time metric $g_{\mu \nu }$
\cite{ref4} 
\begin{eqnarray}
g_{\mu \nu }\rightarrow \tilde g_{\mu \nu }=\exp \left[ -\Phi \right] g_{\mu
\nu }.  \label{conf}
\end{eqnarray}

In the so-called \textit{Einstein frame}, as opposed to the \textit{string
frame} used above, the action (\ref{act1.7}) is

\begin{eqnarray}
S=\frac 1{2\ }\int d^4x\sqrt{-\tilde g}\left[ \tilde R-\frac 12\tilde \nabla
_\mu \Phi \tilde \nabla ^\mu \Phi -\frac 1{12}\exp \left[ -2\Phi \right]
\tilde H_{\mu \nu \rho }\tilde H^{\mu \nu \rho }\right] .  \label{einst-act}
\end{eqnarray}

Since low-energy string theory results in general relativity coupled to the
aforementioned fields in the Einstein frame this frame is often used to find
solutions. However, one should keep in mind that the original string frame
provides the physical interpretation. The classical equations of motion for
a free test string imply that its world-sheet is a minimal surface with
respect to the string frame: the analog of the motion of free point
particles along geodesics. Furthermore, the characteristic size of a
fundamental string is constant in the string frame. So one can describe how,
for example, distances between objects vary w.r.t. their intrinsic size and
it makes sense to talk about 'small' and 'large' curvature scales \cite
{ref18}. Hence, one could argue that strings ``see'' the string frame metric 
$g_{\mu \nu }$ rather than the conformally related Einstein frame metric $%
\tilde g_{\mu \nu }$ \cite{ref5}.

In the Einstein frame the equations of motion are given by \cite{ref4}

\begin{eqnarray}
\tilde R_{\mu \nu }-\frac 12\tilde g_{\mu \nu }\tilde R &=&^{(\Phi )}\tilde
T_{\mu \nu }+^{(H)}\tilde T_{\mu \nu }  \label{e1} \\
\tilde \nabla _\mu \left[ \exp (-2\Phi )\tilde H^{\mu \nu \lambda }\right]
&=&0  \label{H-eq} \\
{\large \tilde \Box }\Phi +\frac 16e^{-2\Phi }\tilde H_{\alpha \beta \gamma
}\tilde H^{\alpha \beta \gamma } &=&0  \label{e3}
\end{eqnarray}
where 
\begin{eqnarray}
\;^{(\Phi )}\tilde T_{\mu \nu } &=&\frac 12(\Phi _{,\mu }\Phi _{,\nu }-\frac
12\tilde g_{\mu \nu }\tilde g^{\alpha \beta }\Phi _{,\alpha }\Phi _{,\beta })
\label{tphi} \\
\;^{(H)}\tilde T_{\mu \nu } &=&\frac 1{12}\left[ e^{-2\Phi }\left( 3\tilde
H_{\mu \lambda \kappa }\tilde H_\nu ^{\;\;\lambda \kappa }-\frac 12\tilde
g_{\mu \nu }\tilde H_{\alpha \beta \gamma }\tilde H^{\alpha \beta \gamma
}\right) \right] .  \label{tH}
\end{eqnarray}
(\ref{H-eq}) can be rewritten using differential forms as 
\begin{eqnarray}
d*\left( e^{-2\Phi }\tilde H\right) =0  \label{astH}
\end{eqnarray}
where $*$ is the Hodge operator. Using its properties and those of the
exterior derivative $d,$ it is found that the antisymmetric tensor field
strength can be expressed in terms of a scalar field $b$: 
\begin{eqnarray}
\tilde H^{\mu \nu \lambda }=e^{2\Phi }\tilde \epsilon ^{\rho \mu \nu \lambda
}b_{,\rho },  \label{b-def}
\end{eqnarray}
where $\tilde \epsilon ^{\rho \mu \nu \lambda }$ is the totally
antisymmetric tensor. Furthermore, there is the Bianchi identity 
\begin{eqnarray}
d\tilde H=0.  \label{hb}
\end{eqnarray}
The dynamics of $b$ are determined by (\ref{hb}), which yields, 
\begin{eqnarray}
{\large \tilde \Box }b+2\tilde \nabla ^\mu \Phi \tilde \nabla _\mu b=0.
\end{eqnarray}

\section{Symmetries}

The four-dimensional low-energy action is invariant under global $O(d,d;\Bbb{%
R})$ and global $SL(2,\Bbb{R})$ transformations, where $d$ is the dimension
of the compactified space.

\subsection{$O(d,d;\Bbb{R})$ Symmetry from Dimensional Reduction}

In \cite{ref10} it is shown that dimensional reduction from $D+d$ to D
dimensions results in a globally $O(d,d;\Bbb{R})$ invariant theory in $D$
dimensions. It is assumed that the $(D+d)$ dimensional space-time can be
written as a direct product of the form $\mathcal{M}\times K$, where $%
\mathcal{M}$ is a $D$ dimensional space-time with local coordinates $x^\mu $ 
$(\mu =0,1,..,D-1)$ and $K$ is a compact ``internal'' space with local
coordinates $y^\alpha $ $(\alpha =1,..,d)$. Furthermore it is assumed that
all fields in the theory are independent of the internal coordinates $%
y^\alpha $. Now consider the action in $D+d$ dimensions 
\begin{eqnarray}
S &=&\int_Ndx\int_Kdy\sqrt{-\hat g}e^{-\hat \phi }(\hat R(\hat g)+\hat
g^{\hat \mu \hat \nu }\partial _{\hat \mu }\hat \phi \partial _{\hat \nu
}\hat \phi -\frac 1{12}\hat H^{\hat \mu \hat \nu \hat \lambda }\hat H_{\hat
\mu \hat \nu \hat \lambda }). \\
\hat \mu ,\hat \nu &=&0,..,D+d  \notag
\end{eqnarray}
The hatted quantities now refer to the $(D+d)$-dimensional space-time. In
order to see that the dimensionally-reduced action is invariant under global 
$O(d,d;\Bbb{R})$ transformations it is important to introduce the following
two quantities. First, the shifted dilaton which is invariant under $O(d,d;%
\Bbb{R})$ transformations, 
\begin{eqnarray}
{\hat \Phi }\equiv \hat \phi -\frac 12\log detG_{\alpha \beta }.
\label{shifteddil}
\end{eqnarray}
Second, the following $2d\times 2d$ matrix $M$, written in $d\times d$
blocks,

\begin{eqnarray}
M\equiv \left( 
\begin{array}{lr}
G^{-1} & -G^{-1}B \\ 
BG^{-1} & G-BG^{-1}B
\end{array}
\right)
\end{eqnarray}
where $B$ denotes the matrix corresponding to the antisymmetric tensor field 
$B_{\alpha \beta },$ while $G$ is that for the internal metric $G_{\alpha
\beta }.$ $M$ is a symmetric $O(d,d)$ matrix. In terms of $M,$ the
low-energy action (\ref{act1.7}) can be written in an $O(d,d;\Bbb{R})$
invariant way; that is, it is invariant under the transformations 
\begin{eqnarray}
\hat \Phi \rightarrow \hat \Phi \;\;\;\;\;\;M\rightarrow \Omega M\Omega ^T
\label{odd tra}
\end{eqnarray}
where $\Omega \in O(d,d;\Bbb{R})$.

Scale-factor duality, which was first described by \cite{ref28}, results
when $G$ and $B$ are assumed to be only time-dependent.

\subsection{$SL(2,\Bbb{R})$ Symmetry}

In addition to the $O(d,d;\Bbb{R})$ symmetry, which is a generalization of
the discrete $O(d,d;\Bbb{Z})$ T-duality group, there is also an $SL(2,\Bbb{R}%
)$ symmetry which is a generalization of the discrete $SL(2,\Bbb{Z})$
S-duality group.

Following \cite{ref31}, the $SL(2,\Bbb{R})$ of the equations of motion
derived from the low-energy effective action can be revealed by first
changing to the Einstein frame. First, we introduce a complex scalar field $%
\lambda $, 
\begin{eqnarray}
\lambda =b+ie^{-\Phi },
\end{eqnarray}
where $b$ is the axionic scalar field defined in (\ref{b-def}). Writing the
equations of motions in terms of $\lambda $ it can be verified that they are
invariant under the transformation 
\begin{eqnarray}
\lambda \rightarrow \lambda ^{\prime }=\frac{\alpha \lambda +\beta }{\gamma
\lambda +\delta }\;\;\;\;\;\;\alpha \delta -\beta \gamma =1
\end{eqnarray}
which is an $SL(2,\Bbb{R})$ transformation for $\alpha ,\beta ,\gamma
,\delta \in \Bbb{R}$.

The strong-weak coupling interpretation of this $SL(2,\Bbb{R})$ symmetry
emerges if the special element with $\alpha =\delta =0$ and $\beta =1$, $%
\gamma =-1$ is considered. Applying this to a pure dilaton solution $\lambda
=ie^{-\Phi }$ results in 
\begin{eqnarray}
\lambda \rightarrow -\frac 1\lambda \Leftrightarrow e^{-\Phi }\rightarrow
e^\Phi .
\end{eqnarray}
Recalling that $e^\Phi $ can be interpreted as the string coupling, we see
that this special transformation exchanges strong and weak coupling. As
shown in \cite{ref31} this symmetry of the equations of motion can also be
extended to the low-energy effective action.

\section{Cosmological Solutions}

Solutions to the low-energy effective action have been found in different
cosmological backgrounds. Friedmann-Robertson-Walker (FRW) backgrounds were
discussed in \cite{ref4}. The role of the antisymmetric tensor field
strength was discussed in \cite{anti}. Inhomogeneous models were also
investigated \cite{inhom}. The solutions usually show an unbounded evolution
of the dilaton, so the string coupling is not bounded. Furthermore, as in
general relativity, curvature singularities are encountered although we
expect higher-order terms to dominate in the vicinity of these ultra-high
curvature regions.

\section{Pre-Big-Bang Cosmology}

The $O(d,d)$ (duality) symmetry inspired a new mechanism for inflationary
evolution which does not rely on the potential energy of a scalar field as
the source for inflation but on the combined action of dilaton and metric 
\cite{ppb}. There are two branches which are related by duality. In the
example discussed below, the (+) branch is the pre-big-bang solution in
which inflation occurs and to lowest order in the inverse string tension $%
\alpha ^{\prime }$ the solution runs into a singularity at some time. The
duality related (--) branch can be smoothly connected to a standard FRW
exhibiting decelerated expansion with a constant dilaton. Then it is assumed
that the dilaton potential becomes important and the dilaton settles down at
its minimum, otherwise there might be problems with observations since a
massless dilaton violates the equivalence principle \cite{ref38} \cite{ref47}%
. To satisfy observational constraints the dilaton mass has to be bounded
below by \cite{ref37} \cite{ref47} 
\begin{equation*}
M_\Phi >10^{-4}eV.
\end{equation*}

In the standard big bang model the beginning of time coincides with the big
bang. The pre-big-bang model dissolves this connection. The beginning of
time is somewhere in the infinite past. The evolution from a weak
coupling/low curvature regime to a high curvature regime provides conditions
for joining the pre-big-bang to the post-big-bang stage in which the
standard cosmological model (nucleosynthesis etc.) applies. This assumes the
solution of the graceful exit problem which all the tree-level inflationary
solutions have to face. In fact, in \cite{ref39} it is shown that a smooth
transition from the (+) to the (--) branch in the low-energy effective
theory (that is, to lowest order in the inverse string tension $\alpha
^{\prime }$), is not possible.

An attractive feature of the pre-big-bang model is that inflationary
solutions follow naturally from the low-energy effective theory. The dilaton
and its dynamics are given by the theory, so there is no need for
assumptions of the kind made in standard inflationary models, where one has
to justify the existence of a scalar field with a certain kind of potential.

A simple example of a pre-big-bang cosmology is set in a flat FRW background
with a string frame metric 
\begin{eqnarray}
ds^2=-dt^2+a^2(t)dx_idx^i\;\;\;\;\;\;i=1,2,3.  \label{met5}
\end{eqnarray}
The dilaton satisfies $\Phi =\Phi (t)$ and for simplicity the antisymmetric
tensor field strength is assumed to be vanishing. The solutions in pre- and
post-big-bang regimes are given by \cite{ref38} \cite{ref39}

\begin{enumerate}
\item  \textit{(+) branch (defined in the domain $t<0$)} 
\begin{eqnarray}
H=\mp \frac 1{\sqrt{3}t}\;\;\;\;\;\;\Phi =\Phi _0+(\mp \sqrt{3}-1)\ln \mid
t\mid   \label{5.12}
\end{eqnarray}

and $a$ evolves as $a=a_0\left( \frac t{t_0}\right) ^{\mp \frac 1{\sqrt{3}}}$%
.

\item  \textit{(--) branch (defined in the domain $t>0$)} 
\begin{eqnarray}
H=\pm \frac 1{\sqrt{3}t}\;\;\;\;\;\;\Phi =\Phi _0+(\pm \sqrt{3}-1)\ln t
\end{eqnarray}

and $a$ evolves as $a=a_0\left( \frac t{t_0}\right) ^{\pm \frac 1{\sqrt{3}}}$%
.
\end{enumerate}

The solutions in the two domains are related by scale-factor duality, $%
a\rightarrow a^{-1}$, and time reflection, $t\rightarrow -t$. The (+) branch
solutions have a future singularity, that is they run \textit{into} a
singularity, whereas the (--) branch solutions have a past singularity, that
is they evolve\textit{\ from} an initial singularity. It should be noted
that these are the defining properties of the (+) and (--) branches. As
pointed out in \cite{ref39}, the identification of the (+) branch with the
domain of negative times and the (--) branch with that of positive times is
just a coincidence. In general, this identification does not necessarily
hold, especially when a dilaton potential is present.

The (--) branch describes either decelerated expansion, that is with $H>0$
and $\ddot a<0$; or accelerated contraction for $H<0$ and $\ddot a>0$.
Choosing the expanding solution it can be shown that it can be smoothly
connected to a standard FRW radiation-dominated solution \cite{ref38}.
Including in the equations of motion a matter source in form of a perfect
fluid with energy density $\rho ,$ it turns out that the FRW
radiation-dominated solution 
\begin{equation*}
a=t^{\frac 12}\;\;\;\;\Phi =const.\;\;\;\;\rho =a^{-4}
\end{equation*}
is a stable solution and it acts as an attractor for the dilaton of the (--)
branch solution \cite{ref38}.

In the (+) branch domain there are solutions describing decelerated
contraction ($H<0$, $\ddot a<0$) and, more interestingly, solutions
describing accelerated expansion ($H>0$, $\ddot a>0$). These latter
solutions describe so called super-inflationary models. Looking at the
evolution of the dilaton in this case it can be seen that the string
coupling $g_s^2$ evolves from very small values as $t\rightarrow -\infty $
to a strong coupling regime on approach to the ``big bang'' as $t\rightarrow
0$, 
\begin{equation*}
g_s^2=e^\Phi \propto \mid t\mid ^{-(\sqrt{3}+1)}.
\end{equation*}

These solutions not only display inflationary behaviour driven by the 
\textit{kinetic} energy of the dilaton but they also evolve from a weak
coupling to a strong coupling regime. This means that the trivial, but
exact, Minkowski solution of string theory could be taken as a starting
point of the cosmological evolution. Furthermore, in the weak coupling
regime the low-energy effective theory used here is perfectly valid.

This simple picture is modified once spatial curvature is admitted. The
authors of ref. \cite{ref41} investigated the impact of spatial curvature on
the FRW pre-big-bang solutions. They found that the pre-big-bang
inflationary solutions with positive or negative spatial curvature (i.e. $k$
positive or negative, respectively) require fine tuning of initial
conditions in order to provide a long enough inflationary period. As
described above, the flat pre-big-bang solutions can be, in principle,
extended infinitely backwards in time while still describing a sensible
cosmological model. However, the solutions with non-zero curvature differ
from the flat ones. Solutions with positive curvature have an initial
singularity. Those with negative curvature have a linearly \textsl{increasing%
} scale factor as $t\rightarrow -\infty $. So in the string frame the
scale-factor starts with a large value at $t\rightarrow -\infty $ and
subsequently decreases to some value $a_{min}$ before increasing up to the
big bang singularity at $t=0$. In ref. \cite{ref41} it was concluded that
this behaviour can delay the onset of inflation and so the inflationary
period might not be sufficient to solve the flatness and horizon problem. In
ref. \cite{ref42} it was argued that the problem with fine tuning in
pre-big-bang inflationary models with spatial curvature is cured by the
introduction of $\alpha ^{\prime }$ corrections.

There are two types of corrections to the low-energy effective action: The
string coupling $g_s^2$ determines the importance of string loop corrections
and the inverse string tension $\alpha ^{\prime }$ controls corrections due
to the finite size of a string. Gasperini et al. \cite{ref17} investigated
the equations of motion in the case when the string coupling is still small,
but $\alpha ^{\prime }$ corrections become important. In this case the
(four-dimensional) action reads \cite{ref17} \cite{ref42} 
\begin{eqnarray}
S=-\frac 1{2\lambda _s^2}\int d^4x\sqrt{-g}e^{-\Phi }\left[ R+(\nabla \Phi
)^2-\frac{\alpha ^{\prime }}4\left( R_{GB}^2-(\nabla \Phi )^4\right) \right]
\end{eqnarray}
where $\lambda _s$ is the string length $\lambda _s^2\propto \alpha ^{\prime
}$ and 
\begin{equation*}
R_{GB}^2\equiv R_{\mu \nu \alpha \beta }^2-4R_{\mu \nu }^2+R^2
\end{equation*}
is the Gauss-Bonnet term.

Considering the spatially-flat case, they found that the dilaton-driven
super-inflationary stage with the scale-factor $a$ behaving as $a\propto
(-t)^{-\frac 1{\sqrt{3}}}$ is attracted to a state of exponential inflation
and a linearly growing dilaton when the first-order $\alpha ^{\prime }$
correction is included \cite{ref17} \cite{ref42}: 
\begin{eqnarray}
a(t)\propto \exp (H_st)\;\;\;\;\;\;\Phi (t)=\Phi _s+c(t-t_s)
\end{eqnarray}
with $H_s$ and $c$ constants. The constant $t_s$ is the time when the
evolution enters the string phase, that is, when $\alpha ^{\prime }$
corrections become important. So in this model there is a super-inflationary
stage followed by a de Sitter stage, thereby regularizing the singularity
which is unavoidable in the low-energy string theory. Maggiore et al. \cite
{ref42} extended the arguments of \cite{ref17} to include spatial curvature
in the dilaton-driven stage and their numerical calculations show that in
all three cases, i.e. for $k=0,\pm 1$, the first-order $\alpha ^{\prime }$
correction regularize the singularity and lead to a stage of exponential
inflation and linearly growing dilaton. They found that this additional
inflationary stage can provide the required amount of inflation while still
satisfying the various observational constraints.

\section{G\"odel Universes}

G\"odel's solution is at once an example of a homogeneous spacetime, a
rotating universe, and an exact mathematical model of a world that contains
closed timelike curves (CTCs). It provides a theoretical laboratory in which
to study the existence and stability of some important global properties of
spacetimes.

We write the general form of the string effective action to $\alpha ^{\prime
}$ order in the string frame as, \cite{firstalfa}, \cite{bdgod}, 
\begin{eqnarray}
S &=&\int d^nx\sqrt{-g}~e^{-\phi }\left\{ R-2\Lambda +(\partial \phi
)^2-\frac 1{12}H^2\right.   \notag \\
&&\ \ \ \ \ \ \ \ -\alpha ^{\prime }\lambda _0\left[ R_{\mu \nu \sigma \rho
}R^{\mu \nu \sigma \rho }-\frac 12R^{\mu \nu \sigma \rho }H_{\mu \nu \alpha
}H_{\sigma \rho }{}^\alpha +\right.   \label{act} \\
&&\ \ \ \ \ \ \ \ \left. \left. \frac 1{24}H_{\mu \nu \lambda }H^\nu
{}_{\rho \alpha }H^{\rho \sigma \lambda }H_\sigma {}^{\mu \alpha }-\frac
18H_{\mu \rho \lambda }H_\nu {}^{\rho \lambda }H^{\mu \sigma \alpha }H^\nu
{}_{\sigma \alpha }\right] +O(\alpha ^{\prime }{}^2)\right\}   \notag
\label{ts}
\end{eqnarray}
where $\lambda _0=-\frac 18$ for heterotic strings, $-\frac 14$ for bosonic
strings and $0$ for superstrings; $n$ is the number of spacetime dimensions, 
$\alpha ^{\prime }$ is the inverse string tension parameter, $\phi $ is the
dilaton, $g$ the determinant of the metric; $\Lambda $ is the cosmological
constant, $R$ the Ricci scalar, $R_{\mu \nu \rho \sigma }$ the Riemann
tensor, and $H_{\mu \nu \rho }$ is the axion with $H^2=H_{\mu \nu \rho
}H^{\mu \nu \rho }$. The cosmological constant term $\Lambda $ is related to
the dimension of space and the inverse string tension by $\Lambda
=(n-26)/3\alpha ^{\prime }.$ This relation holds for bosonic strings; for
superstrings $\Lambda $ is proportional to $n-10$.

Homogeneous spacetime metrics of G\"odel type, \cite{godel}, have the form

\begin{eqnarray}
ds^2=-dt^2-2C(r)dtd\psi +G(r)d\psi ^2+dr^2+dz^2,  \label{line}
\end{eqnarray}
where the functions $C(r)$ and $G(r)$ have the forms 
\begin{eqnarray}
C(r) &=&\frac{4\Omega }{m^2}\sinh ^2{\left( \frac{mr}2\right) },  \notag \\
G(r) &=&\frac 4{m^2}\sinh ^2{\left( \frac{mr}2\right) }\left[ 1+\left( 1-%
\frac{4\Omega ^2}{m^2}\right) \sinh ^2{\left( \frac{mr}2\right) }\right] ,
\label{G1}
\end{eqnarray}
with $m$ and $\Omega $ constants, \cite{tiomno}. In order to\textit{\ avoid }%
the existence of CTCs in these spacetimes we require the \textit{no
time-travel condition} $G(r)>0.$ This is consistent with the \textit{%
chronology protection conjecture} of Hawking, \cite{hawk}.

G\"odel universes rotate: the four-velocity of matter is $u^\alpha =\delta
_0^\alpha ,$ the rotation vector is $V^\alpha =\Omega \delta _0^3,$ and the
vorticity scalar is given by $\omega =\Omega /\sqrt{2}$. The original
G\"odel metric of general relativity, \cite{godel}, has $m^2=2\Omega ^2$ and
clearly violates the no-time-travel condition. There has been extensive
discussion about the generality and significance of the presence of CTCs in
the G\"odel metric in general relativity (see refs.\cite{nahin},\cite{metric}%
).

The only nonvanishing components of the Riemann tensor in an orthonormal
frame permitted by the spacetime homogeneity of the G\"odel universe are
constant, \cite{accio}, with $R_{0101}=R_{0202}=\Omega ^2,R_{1212}=3\Omega
^2-m^2.$ The dilaton depends only on the coordinate along the axis of
rotation, $z$, so $\phi =\phi (z)=fz+\phi _0,$where $f$ and $\phi _0$ are
constants and $H_{012}=-H^{012}=E,$ with $E$ constant. The field equations
for the G\"odel metric now reduce to the three polynomials.  We now have
three equations and six constants, leaving three (say, $\Lambda $, $f$ and $E
$) arbitrary. We obtain $\Omega ^2=m^2/4\ $which confirms the existence of a
G\"odel solution which fulfils the no time-travel condition. The value of $%
\alpha ^{\prime }$ can now be expressed in terms of the velocity of rotation
of the universe, $\Omega ,$ and the strength of axion field, $E$, giving 
\cite{bdgod}, 
\begin{equation}
\alpha ^{\prime }=\frac{E^2-4\Omega ^2}{4\lambda _0\left( 4\Omega
^2+E^2\right) (\Omega ^2-\frac 54E^2)}.
\end{equation}
These are string theoretic generalisations of the famous rotating G\"odel
universes of general relativity. However, unlike in general relativity,
string solutions exist which forbid time travel. These new solutions display
remarkably simple relations between the rotation of the universe and the
string tension by virtue of their $T-$duality symmetry. The simplicity of
these relationships suggests that the remarkable symmetries of string theory
may constrain peculiarities in the causal structure of spacetime that
general relativity permits.

\textbf{Acknowledgements}

The authors would like to thank E. Copeland, M. D\c abrowski and J. Lidsey
for discussions. JDB is supported by PPARC and KEK was supported 
by the German National Scholarship Foundation. 

$\ $\\

\end{document}